# Open Data Visualization:
# Keeping Traces of the Exploration Process

Benoît Otjacques, Mickaël Stefas, Maël Cornil, Fernand Feltz
Public Research Centre - Gabriel Lippmann
41, Rue du Brill, L-4422 Belvaux, Luxembourg
{otjacque, stefas, cornil, feltz}@lippmann.lu

## ABSTRACT
This paper describes a system to support the visual exploration of Open Data. During his/her interactive experience with the graphics, the user can easily store the current complete state of the visualization application (called a viewpoint). Next, he/she can compose sequences of these viewpoints (called scenarios) that can easily be reloaded. This feature allows to keep traces of a former exploration process, which can be useful in single user (to support investigation carried out in multiple sessions) as well as in collaborative setting (to share points of interest identified in the data set).

## Categories and Subject Descriptors
H.5.2 [**Information Interfaces and Presentation**]: User Interfaces - *Graphical User Interfaces.*

## General Terms
Human Factors.

## 1. INTRODUCTION
Several Governments worldwide have recently given free access to large amounts of data about very diverse topics like economics, mobility or environment. This movement is usually called Open (Governmental) Data. However, like many other authors, Agrawala et al [1] point out that acquiring and storing data is by itself of little value. Visualization is one of the answers to transform them into knowledge providing that efficient graphics are used. Many of the numerous systems offering interactive visualizations can be used to display Open Data. In some cases, the interaction between the user and the advanced visualization may really be as immersive as games, which gives a very rewarding experience. Unfortunately, this may also leave the user with an imprecise feeling of data understanding. It may be impossible for him/her to locate again what he/she has discovered, for instance to communicate or share a specific aspect of the data. This paper presents how we have added a new feature to an existing visualization tool in order to help the user to easily store, group and recall specific points of interest encountered during the data exploration process.

## 2. STATE-OF-ART
In recent months, an increasing number of Governments worldwide (e.g. USA [15], UK [14], France [12]) have decided to give access to some of the data they produce or manage. Recent examples include Spanish Census Data [4], US health statistics [10] as well as some initiatives in Kenya [13].

This increasing amount of data is made available to be explored and this will probably change how citizens and other stakeholders (e.g. press, lobbyists, NGOs...) deal with public affairs. From this perspective, Yuille and Macdonald [11] stress the fact that since Open Data aims to promote increased transparency it is of utmost importance to use technology with a low entry barrier.

Yuille and Macdonald [11] have designed a framework relying on visualizations to explore Open Data and to share the related findings with others (cf. idea of shared storytelling). They position their proposal as an object-centered social network (like e.g. Flickr) by opposition to ego-centered ones (like e.g. Twitter). In this context graphics are the objects at the center of the social network. To make storytelling occur, these authors explain that the visualizations need an identity. This helps people to recognize them, to locate them and to assess whether they want to comment them. They named "*decoration*" the process of giving such an identity in the network to a visualization. Among others, they also claim that annotation features must be very simple. Furthermore, the process must create the annotated visualizations as derivative ones keeping intact the original.

A multi-step process is often needed to progressively design efficient visualizations. Scheidegger et al. [6] have highlighted the advantages of keeping trace of this history. They showed that this knowledge collected from multiple processes may be reused to help the user to construct new visualizations by analogy with previous design or exploration pipelines. Keeping traces is also helpful in the context of data exploration processes that include numerous steps where the user interact with various data views. VisTrails [16] is an example of a tool combining data exploration and visualization modules as well as a recording system "*maintaining detailed history information about the steps followed and data derived in the course of an exploratory task*".

The visual exploration of Open Data also requires easy means to store some points in the process where meaningful elements were discovered or confirmed. Coming back easily to some of these points of interest should be supported to avoid the effort to reconfigure the visualization tool exactly as it was at that point of the exploration process (which may take a considerable amount of time if the number of parameters is high). In some cases, some points of interest differ little from each other and the user may face difficulties to be reminded what precise point he/she wants to go to. Even when the graphics are displayed, it may be difficult to identify exactly which parameters have been modified between them. Although it is not focused on visualization software, a similar motivation has lead Drucker et al. [3] to propose a system to compare, manage and reuse slide presentations. From a broader point of view several techniques have been suggested to manage the history of documents by visually highlighting the differences and the similarities among them (e.g. [9], [2]). Similar work is also needed for visualizing the differences among graphics.

Numerous authors have pointed out the difficulty to design efficient graphics (e.g. [1]). Hullman et al. [5] recently drew attention to another point: the influence of social information on quantitative judgments in graphical perception. With the emergence of collaborative visualization platforms (like Many Eyes [8]), we must keep in mind the risk of cascade misinterpretations or cumulative adoption process of specific charts depending on first users (possibly wrong) choice. Indeed, Hullman et al. [5] have showed with an experiment that presenting social information (i.e. other people's prior reactions about the same chart) influences the judgment of someone seeing a chart. This aspect should be taken into consideration when designing multi-user systems such as those dedicated to visualize Open Data.

To sum up, prior research has pointed out several elements to take into account to visualize Open Data: (1) technology must have a low entry barrier; (2) "decoration" may help to support story telling around visualizations; (3) keeping traces of history is needed; (4) visual representation of differences may help to manage a collection of similar graphics; (5) social elements influence the interpretation of graphics.

## 3. RESEARCH RATIONALE

In former projects our team has progressively developed a innovative visualization tool called Calluna. This software has been tested in several contexts (e.g. finance industry and Human Resources management) with promising results. Recently we have explored its potential to visualize Open Data.

However, we have observed that some users face difficulties due to the increasing number of features and options available in the tool. Some pilot users explicitly asked us to add a simplifying layer on the tool. If this increasing complexity may (at least partly) be lowered with tutorial sessions and rework of the User Interface, we have started to study others solutions. A specific problem was caused by the difficulty to replay an exploratory process of a dataset. Since the users can easily load several datasets, switch from one data view to another, filter data interactively and change display preferences (e.g. colors), they can dive into the data and live an immersive experience. Unfortunately, without a trace of this history, it is almost impossible for the user to replay this process or to locate a specific point of interest in the dataset. This limitation may result in frustrating attitudes and rejection of the visualization tool. For illustrative purposes, the last version of our visualization software offers several hundreds of options to the user to configure numerous aspects of the tool. It is clearly beyond any user capability to remember the exact configuration at a given moment of time.

This issue is also critical in collaborative contexts involving several data analysts. In this paper we focus on asynchronous collaboration. Indeed, we study how to keep trace of the exploration history and to be able to investigate it again in a individual or collective setting, which clearly refers to an asynchronous process. Moreover, some solutions have already been proposed to support synchronous collaboration where people distributed in various locations see and interact simultaneously with the same visualizations.

Typically, John is a journalist analyzing Open Data about public finances (cf. Figure 1). During the exploration of these data he identifies two valuable graphics: one highlighting a possible correlation between public investment in education and the unemployment rate in a given region (G3) and a second one confirming his prior hypothesis regarding non justified expenses in a given ministry (G5). He wants to share G5 with his colleagues to prepare a newspaper article and to take advice about G3 from Helen who is an expert of the job market. The simplest (and probably most often used) solution for him is to generate screenshots or images of the graphics and to send them by usual means (e.g. e-mails). Unfortunately, this approach provides limited potential for further analysis to the people receiving them. A better option would be to send them some means to put them exactly in the same configuration of the visualization tool where the valuable graphics G3 and G5 were identified. For instance, Helen could directly start from G3 and explore the data relying on her own experience and expertise. She could for example find another graphics G9 illustrating better whether the correlation found by John is statistically relevant or not. In this case, she would benefit on the one hand from John's initial work (which helps her to locate the interesting starting point G3) and on the other hand from the whole interaction features of the visualization tool (which allows her to reach G9).

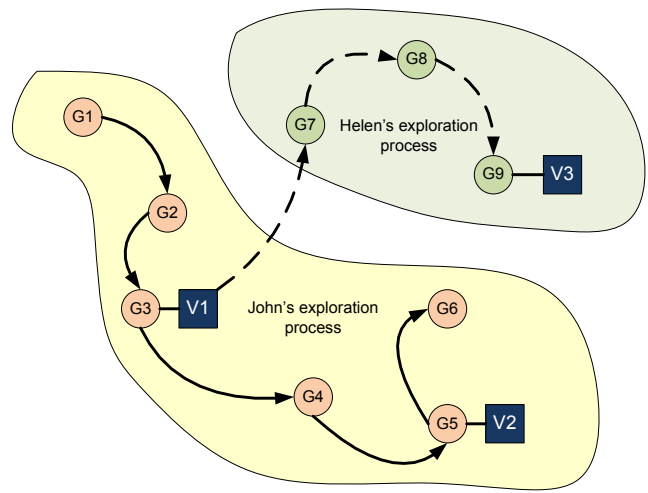

**Figure 1: Data Exploration Process**

Therefore, we decided to add new features supporting the user to come back to specific configurations of the application and to navigate easily from one point of interest to another without requiring the complete manual reconfiguration of the visualization environment. This extension of the tool was also justified by prior research findings discussed in the state-of-art section.

## 4. CONCEPTUAL MODEL

Before discussing our conceptual model to support the data exploration process, we need first to describe some fundamental elements of our visualization software.

Several datasets coming from various sources (e.g. database, file server, web service...) may be simultaneously loaded in the tool. Each dataset may be visualized with several distinct *views* (e.g. linked with a Master-Detail relationship). Basically a *view* is a type of graphics (e.g. pie chart, treemap, table...). A dataset is called a *relation* to recall that it can be displayed as a set of related *views*. Each *view* has a reference data item called *current node* which allows to specify which data subset is displayed in this view or in the related ones (cf. Master-Detail relationship). Each *view* displays a specific subset of the data *attributes* (and if applicable for given period(s) of time). Thanks to *filters*, the data items displayed in a *view* may be highlighted according to the satisfaction of criteria related to the *attributes*.

Now we can describe the conceptual model previously mentioned. It relies on two basic concepts: viewpoints and scenarios. A **Viewpoint** can be described as a point of interest or a milestone in the data exploration process (e.g. V1, V2, V3 in Figure 1). A viewpoint specification includes the complete configuration of the application as well as the data views currently displayed at a given moment of time. It is comparable to an interactive screenshot or an enhanced bookmark. Indeed, it does not only provide a static image (cf. usual screenshot) or a simple link to a resource (cf. bookmark) but the complete state of a complex environment to visualize data. A viewpoint can help a user to come back to where he/she or another user (in a collaborative asynchronous setting) found meaningful information. A **scenario** is defined as a sequence of viewpoints (cf. Figure 2). It can be used to support and keep traces of the data exploration process taking into account its temporal evolution. A scenario allows the user to navigate among the viewpoints without passing through every step of the initial process. Of course, the user can start exploring the data with all features of the visualization tool from every viewpoint of a scenario. In other words, a scenario is a suggested path for exploring the data and not a mandatory procedure.

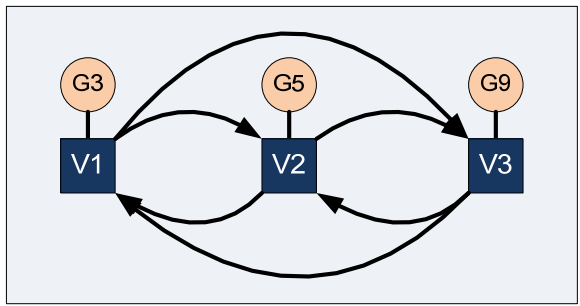

**Figure 2: Scenario**

Keeping traces of the data exploration process includes several aspects. In our visualization tool, these aspects are managed by *preferences* that include the configuration variables of the tool and the variables describing the current state of the user interface (e.g. data displayed, layout, active filters...). Some preferences are explicitly set by the user via a dedicated menu (e.g. *background color in pie chart*). Others are implicitly set during the interactive exploration of the data (e.g. *current node*, *attributes displayed*, *filter limits*). Preferences are the building blocks of viewpoints and scenarios. Indeed, a viewpoint is simply the complete list of preferences and their value at a given moment of the exploration process. The basic principle is that loading several times the same list (i.e. a viewpoint) will bring the visualization tool into the same configuration. The user will see the same data, visualized by the same graphics showing the same attributes, potentially highlighting the same data items on the basis of the same filters... Consequently, all preferences must be stored to be able to reload the tool exactly as it was at a given point of the data exploration process.

In our visualization tool the preferences are grouped according to two dimensions: the **scope** and the **nature** of their influence. Table 1 illustrates this two-dimensional model.

**Table 1. Preferences Model**

| Nature | | Scope | | |
|---|---|---|---|---|
| | | Application | Relation | View |
| Category | Pref. | | | |
| Cat 1 | $P_{1,1}$ | Yes | No | No |
| Cat 2 | $P_{2,1}$ | No | Yes | Yes |

The user may use the preferences with three levels of **scope**: the complete application, a given *relation* or a given *view*. Some preferences can be set at every level while others only apply to one or two of them. Consequently, the set of preferences available at each level of scope is different. For instance, the preference $P_{1,1}$ only applies to the application level (e.g. *default number of Master and Details views for a dataset that has never been loaded before*). As a second example, the preference $P_{2,1}$ can be set at the relation or the view level (e.g. *choice of attributes associated to the axes in a temporal chart*).

The *preferences* may influence various aspects (subsequently called categories) (cf. **nature** dimension). Some examples of categories are listed below:

- user interface global layout,
  e.g. *default number of Master and Details views and their respective position at initialization*

- data displayed,
  e.g. *attribute(s) associated to axes in a 2D chart*

- view-specific aspects,
  e.g. *choosing whether the background color of slices in a pie chart convey some information or whether they are simply chosen to be easily distinguishable*

- filter-specific aspects,
  e.g. *choice of the scale used to display how much the data items satisfy a combination of filter criteria*

- timeline configuration,
  e.g. *maximum number of distinct periods of time that may be simultaneously managed by the timeline*

- import / export features,
  e.g. *default path for the import feature for a given file format (like CSV)*

- localization features,
  e.g. *format of numbers*

Obviously, the preferences differently influence the specification of a viewpoint. For instance, changing the relation (i.e. changing the data set) is more important than modifying the size of the window displaying it. This aspect will be taken into account when we will discuss how much a given viewpoint differs from another.

## 5. IMPLEMENTATION

This section explains how we have implemented the conceptual model into our visualization tool. We have identified three basic features to be offered to the user:

1. to save and reload viewpoints,
2. to group some viewpoints in a scenario,
3. to visualize the difference between some viewpoints.

They are discussed in the following subsections.

### 5.1 Save and load viewpoints

Saving and loading viewpoints are the first features that we must provide because the two others directly depend on it. We have explained earlier that a viewpoint is a list of preferences and their value. The user can save and load a viewpoint whenever he wants via a dedicated menu. These features conform to the ubiquitous paradigm "*Open/Save file*", which makes it easy to understand. In fact, a viewpoint is implemented as an XML file managed with "*Open/Edit/Save Viewpoint*" menu items.

In addition to the preferences, a viewpoint file also includes three types of metadata to help the user to know or to remember what it refers to: (1) metadata about the **file** itself, (2) semantics metadata about the **content** of the viewpoint and (3) opinion metadata expressed by the file **owner**. These metadata add "decoration" to the viewpoint (cf. [11]). We think that this feature offers added value to the user because it is very difficult to summarize the description of a viewpoint simply and concisely. Indeed, the basic design rule for our visualization software is to keep the user interface as simple and intuitive as possible. The consequence is that the user does not realize the complexity of the internal model of the application (and he/she does not need to do so). We think then that we must give him/her some clues to manage the viewpoints easily without being forced to learn every detail about them. This is the basic motivation to add metadata to the viewpoint files. To lower the entry barrier of this feature, we have tried to find a good trade-off between manually encoded and automatically derived metadata. All of them are displayed in the "*Save viewpoint*", "*Edit viewpoint*" and "*Open viewpoint*" windows (cf. Figure 3).

The **file** metadata regroups the usual information for managing files: the file name, the place when it is/will be saved (path), the timestamp of the last save operation. This is obviously not original but still useful because the user can rely on his previous knowledge to locate the viewpoints just like another file. Furthermore, we allow the user to link a viewpoint to an image. This picture can be, for instance an icon illustrating the viewpoint topic, a screenshot of the visualization application when the viewpoint was created or whatever else that "speaks to" the user.

Regarding the viewpoint **content**, we consider time and space to be fundamental cognitive anchors, especially for Open Data. Therefore, we allow the user to choose the geographical area that the viewpoint concerns (e.g. a city, a country, the world) in a preloaded list of names and illustrative icons (e.g. flag for a country). Second, if temporal data is loaded in the tool, the period of time that is visualized is also critical to define the current viewpoint. For instance, which fiscal year does this data refers to? This information is automatically derived from the application internal variables and encoded in the viewpoint XML file. Finally, a viewpoint also has a classic "*Description*" field.

Knowing who has saved or edited (subsequently called **owner**) a given viewpoint also helps to get a right overview about it. Our application automatically encodes the owner's name (derived from the login information of the active session on the computer) of a viewpoint as metadata. We have also decided to include some global information about the owner's opinion because it may be valuable for himself (feelings often plays a role in remembering things) but also for others (to encourage discussion about a given topic).

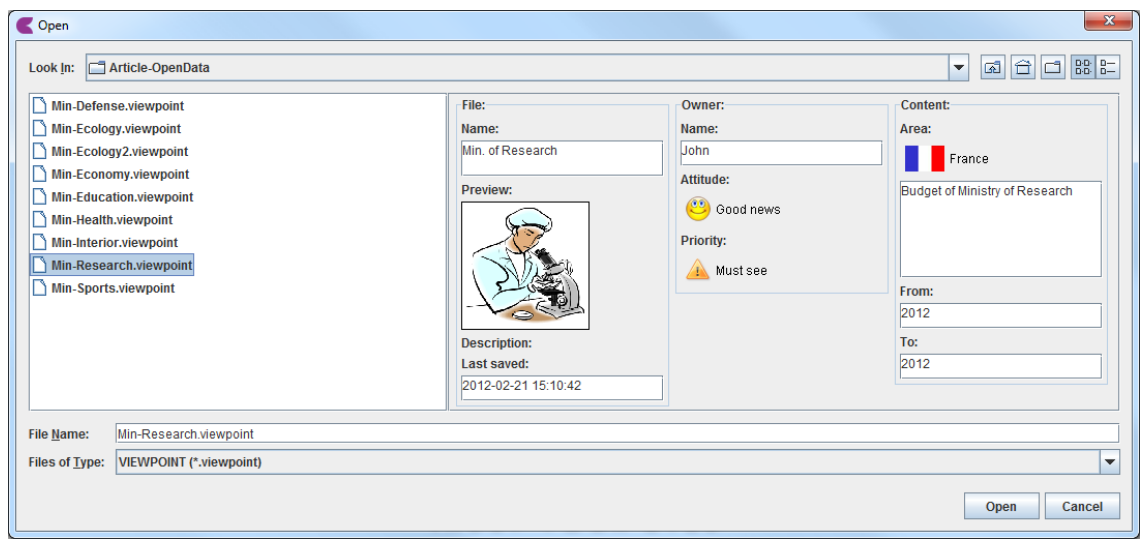

**Figure 3: Open Viewpoint window**

Nevertheless, considering biases that may be caused by prior exploration of the data by other users (cf. [5]), we have limited the number of opinion metadata (that are subjective by definition). First, the priority (*Must see - Interesting - Facultative*) helps to assess the importance of the viewpoint for its owner. Note that viewpoints that are not relevant or that do not highlight any valuable aspect of the data should not exist since viewpoints are worth being saved by definition. Second, the attitude (*Good news - Neutral - Bad news*) informs about how positive/negative the viewpoint is considered by its owner.

## 5.2 Group viewpoints in scenarios

A scenario is simply an ordered list of viewpoints. It is implemented as an XML file including the references to the related XML viewpoint files.

Theoretically, scenarios (like viewpoints) should be given some "decoration". However, they already include indirectly the metadata about the viewpoints (included in the viewpoints XML files). Therefore, only two elements are added at the scenario level: the name and location of the XML file of the scenario.

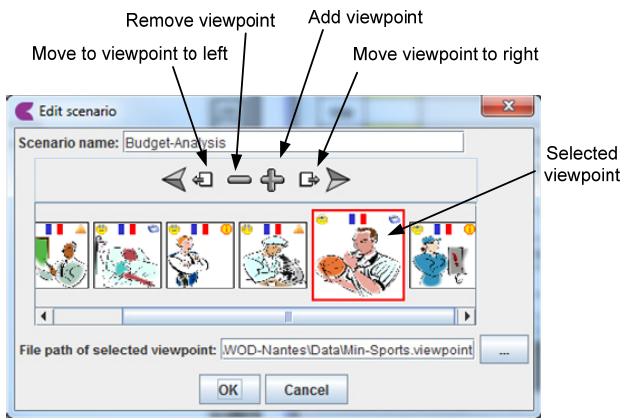

**Figure 4: Edit Scenario**

To create or edit a new scenario, the user uses a dedicated window (cf. Figure 4). He/she can select some viewpoints in the file system, add them at the appropriate place or move them to rearrange the sequence. A preview zone displays the images associated to the viewpoints to help the user to remember what they concern. Some icons illustrating metadata considered to be important are also added above the images (e.g. flag of the area concerned, smiley for owner's attitude). Moreover, a tooltip displays all metadata of the viewpoint when the mouse cursor is placed over the image. The "*Open Scenario*" window also includes a similar component providing an overview of the scenario before loading it into the application.

## 5.3 Visualize differences between viewpoints

In order to visualize what differs between two viewpoints, we first need to identify the differences between them. From a technology perspective, this means identifying the differences between two XML files. The result is another XML file listing all preferences (grouped by category) whose values differs in both viewpoints.

We have already stressed the fact that the global distance (i.e. a metric of differences) between two viewpoints V1 and V2 should take differently into account the various preferences. Each preference is therefore given a weight according to the magnitude of its influence on the viewpoint. The distance between V1 and V2 is computed as the weighted sum of all preferences that differs between them. At this stage of our research, we have simply given a weight to each preference according to our experience with our tool. However, in future steps, we plan to investigate how to refine this model. Our work has some similarities with Shrinivasan and van Wijk's research [7] about the comparison of visualization's states. They proposed to describe the user exploration with four key aspects in which visualization states and user's actions are sorted. However, they used Venn diagrams to display the similarity for each of the four key aspects.

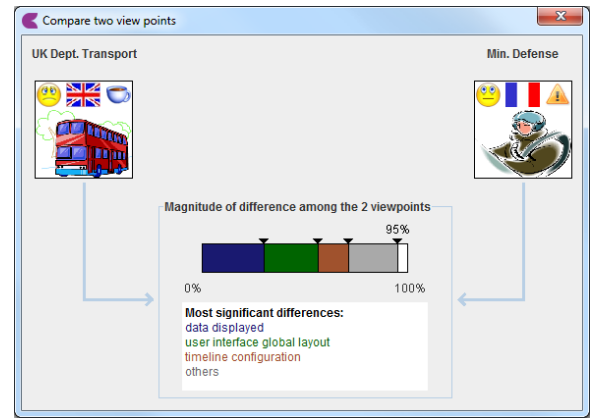

**Figure 5: Comparison of two very different viewpoints**

We can illustrate our approach with data from the French Open Data web site regarding the budget of all Ministries. The user can load this dataset and create several viewpoints for each ministry. We can also load another file about all primary schools to point out some issues in the Ministry of Education. A complex analysis scenario can then progressively be built. In this example, it may be useful for the user to know how close a given viewpoint is to another. A specific window allows to compare two viewpoints (cf. Figure 5). The distance between them is displayed on a colored bar. While the low limit of the scale (i.e. distance = 0 corresponds to 0%) is trivial to compute, the upper limit (i.e. which distance corresponds to 100%?) is more difficult to assess. Therefore, we have adopted an heuristics-based rule derived from tests carried out with various viewpoints to map the right limit of the color scale (i.e. completely different viewpoints) to a computed distance between the two viewpoints. In addition to the global magnitude of difference, we also visualize the three categories of preferences with the highest computed distances among them. This information helps the user to know what differs most in both viewpoints. For instance, Figure 5 illustrates two very different viewpoints (global difference = 95%). In fact, they mainly differ by the data displayed, the user interface global layout and the timeline configuration.

We have also investigated how to visually compare n>2 viewpoints. We have used the classic Multi Dimensional Scaling (MDS) method to position the points associated to the viewpoints on a 2D space (Figure 6). Links can be drawn between any pair of points to display the distance between them. To deal with well known limitations of projection methods like MDS we have added some information about the projection quality, which is intended to help the user to assess its reliability. The label of the links can display the computed distance (before the MDS projection), the layout distance (after the MDS projection) or the ratio of both distances. The mean, variance and distribution of this ratio is also provided (see Projection Quality Metrics frame in Figure 6). We have also added a feature to define a scenario simply by drawing a graph in the 2D view. The user can directly run this new scenario from the 2D view to successively load the related viewpoints in the visualization software.

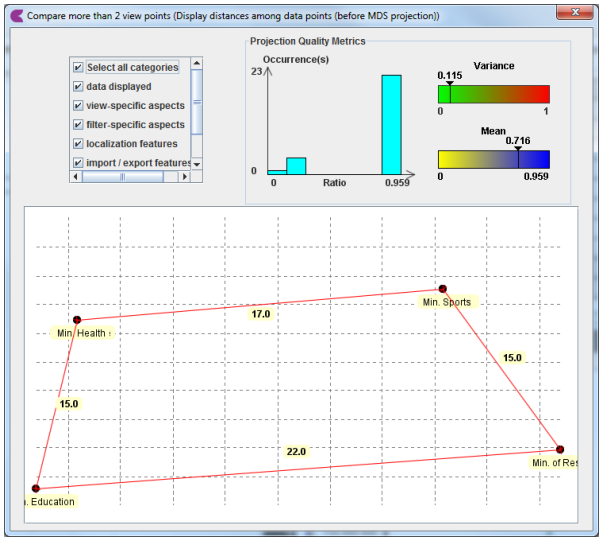

**Figure 6: Comparison of n viewpoints**

## 6. CONCLUSIONS

The exploration process of Open Data is often complex and distributed among several persons during disjoint periods of time. Some advanced visualization software can lead the user to really dive into the data. He/she freely explores various facets of the data, tries to draw relationships among some items, modifies the graphics interactively... This process of gaining insight is one of the added values of visualization software compared to static graphics. Unfortunately few systems allow to keep traces of the history of this visual investigation although it is a key element to remember or share knowledge about the dataset. This paper describes a new module added to an existing visualization tool to tackle this issue. Thanks to the concept of viewpoints and scenarios the user can easily store, share and reload the state of our visualization application exactly as it was when valuable knowledge was discovered.

Furthermore, we propose a global metrics to compare viewpoints computed from all aspects influencing the visualization tool configuration at a given moment. The current model of distance will be refined in further steps of our research. In the future we also plan to evaluate the concept of viewpoints and scenarios with pilot users, especially in a collaborative setting. We hypothesize that some viewpoints coming from former exploration processes will be reused as starting points of new data investigations but we have not assessed this assertion yet.